\documentstyle[aps,amstex,amssymb,preprint,epsf]{revtex}

\tightenlines

\begin{document}                                                       
   
\draft
                                            
\title {Long distance tunneling}

\author{Boris Ivlev} 

\address
{Department of Physics and Astronomy and NanoCenter\\
University of South Carolina, Columbia, SC 29208\\
and\\
Instituto de F\'{\i}sica, Universidad Aut\'onoma de San Luis Potos\'{\i}\\
San Luis Potos\'{\i}, S. L. P. 78000 Mexico}

\maketitle

\begin{abstract}
Quantum tunneling between two potential wells in a magnetic field can be strongly increased when the potential barrier varies in the 
direction perpendicular to the line connecting the two wells and remains constant along this line. A periodic structure of the wave 
function is formed in the direction joining the wells. The resulting motion can be coherent like motion in a conventional narrow band
periodic structure. A particle penetrates the barrier over a long distance which strongly contrasts to WKB-like tunneling. The whole
problem is stationary. The coherent process can be influenced by dissipation.

\end{abstract} \vskip 1.0cm
   
 
\narrowtext

Can a particle move under a long and almost classical potential barrier where classical motion is impossible?

According to Wentzel, Kramers, and Brillouin (WKB) \cite{LANDAU}, there is a finite probability $w\sim\exp(-A)$ of quantum tunneling 
through an one-dimensional potential barrier. This probability becomes negligible for semiclassical barriers when $A=2{\rm Im}S/\hbar$ and 
the classical under-barrier action ${\rm Im}S$ is big. In two-dimensions the most convenient way to calculate the exponent $A$ is by use of
a classical trajectory $x(\tau)$, $y(\tau)$ in imaginary time $t=i\tau$ \cite{COLEMAN1,COLEMAN2,MILLER,SCHMID1,SCHMID2}. The trajectory 
goes in a classically forbidden area (under the barrier) and connects two classically allowed regions having zero velocities at the borders 
of classical regions. The classical action, constructed by means of this trajectory, is called Euclidean action and determines WKB-type 
exponent $A$. The method of classical trajectories in imaginary time is powerful and relatively simple since it allows to determine a 
tunneling probability in the main (no pre-exponent) approximation $\exp(-A)$ just only solving Newton's equation of motion. 

The problem of quantum tunneling in magnetic field was addressed in Refs.\cite{SHKL1,SHKL2,THOUL}. In the Landau gauge there is a parabolic 
gauge potential $m\omega^{2}_{c}(x-x_{0})^{2}/2$ superimposed upon the tunnel barrier potential. The cyclotron frequency is 
$\omega_{c}=eH/mc$ and tunneling occurs in the $x$-direction. If the tunnel barrier is not a constant, containing weak impurity centers, 
$x_{0}$ becomes spatially dependent resulting in a variable gauge potential of a sawtooth shape instead of pure parabolic one. This 
potential is ``pinned'' by impurities, separated by the characteristic distance $b$, and repeats their positions \cite{SHKL1,SHKL2}. In the
regime of a strong magnetic field, when the energy $m\omega^{2}_{c}b^{2}/2$ exceeds a height of the tunnel barrier, an electron tunnels 
incoherently through each peak of the gauge potential. For this incoherent motion the total probability of tunneling is a product of partial
ones.

This process can be elegantly described in terms of classical trajectory in imaginary time $(t=i\tau)$ \cite{GESH} (see also 
\cite{BLATT,GOROKH}). The classical trajectory is constructed in the following way: $x(\tau)$ is real and corresponds to a translational 
motion in the direction of tunneling ($x$-axis) but the transverse coordinate $y=i\eta(\tau)$ is imaginary and performs oscillations in 
$\tau$. The oscillations are produced by an oscillatory force in the $x$-direction since impurities are distributed along the $x$-axis. 
This potential force is balanced by the $x$-component of the Lorentz force $m\omega_{c}\partial \eta/\partial\tau$. The tunnel potential 
$U(x,y)$ is supposed to be even with respect to $y$ and the classical trajectory satisfies the relation 
$U\left[x(\tau),i\eta(\tau)\right]=E$. At the ends of the trajectory the relation $\eta =0$ holds which allows to match with the physical 
$y=0$.

If the magnetic field is not big, kinetic energy enters the game and a scenario of tunneling can be dramatically different. We consider 
the two-dimensional tunnel potential which in the barrier region has the form $U(x,y)=u(y)$, tunneling occurs in the $x$-direction, and 
the magnetic field is aligned along the $z$-axis. In this case a variable gauge potential can form under the barrier a certain structure 
periodic in the direction of tunneling with a period $\Delta x$. An electron moves in this periodic potential in the way similar to 
conventional motion in a periodic structure with a narrow energy band $\Delta E$ where tunneling processes through subsequent periodic
barriers are strongly coherent. For a conventional periodic structure a wave packet can pass over a long distance with no exponential 
decrease in amplitude but only with a delay time $\hbar/\Delta E$ as a velocity in a narrow band is proportional to $\Delta E$ \cite{ZIMAN}. 
In our case, since the potential is $x$-independent in its barrier part, a period in the $x$-direction 
$\Delta x=\sqrt{2|E|/m}\hspace{0.1cm}\Delta\tau$ ($\sqrt{2|E|/m}$ plays a role of velocity) has an intrinsic nature. As shown below, the 
time $\Delta\tau$ is a period of oscillations in the well associated with the potential 
\begin{equation} 
\label{1}
v(\eta)=u(i\eta)-\frac{m\omega^{2}_{c}}{2}\left(\eta +\eta_{0}\right)^{2}
\end{equation}
The particle energy $E$ is negative, $\eta_{0}=\sqrt{2|E|/m\omega^{2}_{c}}$, and the minimum of $u(y)$ corresponds to $u(0)=0$. The 
transverse oscillatory motion in the direction of $y=i\eta$ is coupled to the translational motion ($x$-direction) due to the Lorentz 
force. 

When the magnetic field is close to the certain value $H_{R}$ the probability of tunneling $w\sim\exp(-A)$ through a long barrier becomes 
not exponentially small, like for a conventional narrow band dynamics, and corresponds to $A\rightarrow 0$. This is a situation of 
Euclidean resonance studied in papers \cite{IVLEV1,IVLEV2,IVLEV3,IVLEV4,IVLEV5} for tunneling through nonstationary barriers when also 
$A\rightarrow 0$ at a certain value of an ac amplitude. Therefore, a phenomenon of Euclidean resonance has rather a general nature since 
it occurs also in a static barrier. As argued in this paper, the magnetic field sets a long distance under-barrier coherence which allows 
under-barrier motion over a long distance. This strongly contrasts to WKB-like tunneling.

\begin{figure}[p]
\begin{center}
\vspace{1.5cm}
\leavevmode
\epsfxsize=\hsize
\epsfxsize=12cm
\epsfbox{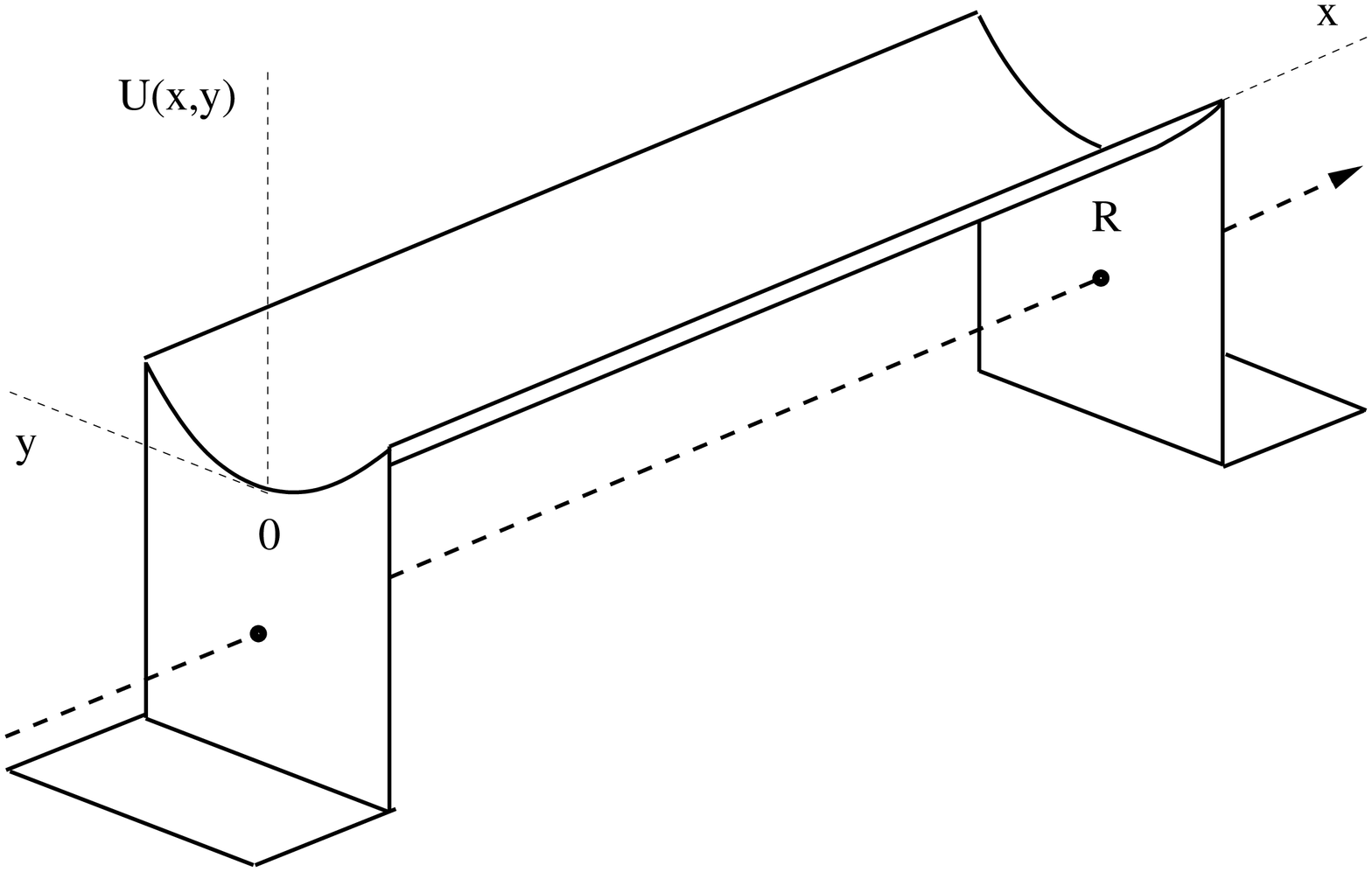}
\vspace{1cm}
\caption{A particle passes (the dashed line) through the tunnel potential.}
\label{fig1}
\end{center}
\end{figure}

Below we formulate the above arguments in terms of a classical trajectory in imaginary time $(t=i\tau)$. A potential barrier, part of 
which is plotted in Fig.~\ref{fig1}, is even with respect to $y$, it does not depend on $x$ at $0<x<R$ where $U\left(x,y\right)=u(y)$ 
($u(0)=0$), and the function $U\left(x,0\right)$ has jumps at $x=0$ and $x=R$. Tunneling occurs between two classically allowed regions 
$x<0$ and $R<x$. For convenience, the potential in  Fig.~\ref{fig1} is drawn in a way that it is a constant at $x<0$ and $R<x$. This 
condition is not necessary and $U\left(x,y\right)$ can correspond to tunneling between two quantum wires or two quantum dots. In order 
to calculate a tunneling probability in the exponential approximation one can know only a classical trajectory in imaginary time 
connecting two points $\{x=0,y=0\}$ at the moment $\tau=\tau_{0}$ and $\{x=R,y=0\}$ at the moment $\tau=0$. Classical equation of motion 
under the barrier have the form  
\begin{equation} 
\label{2}
m\hspace{0.1cm}\frac{\partial ^{2}x}{\partial\tau^{2}}+m\omega_{c}\hspace{0.1cm}\frac{\partial\eta}{\partial\tau}=0;\hspace{1cm}
m\hspace{0.1cm}\frac{\partial ^{2}\eta}{\partial\tau^{2}}+m\omega_{c}\hspace{0.1cm}\frac{\partial x}{\partial\tau}+
\frac{\partial u(i\eta)}{\partial\eta}=0
\end{equation}
The total energy conserves
\begin{equation} 
\label{3}
E=-\frac{m}{2}\left(\frac{\partial x}{\partial\tau}\right)^{2}+\frac{m}{2}\left(\frac{\partial\eta}{\partial\tau}\right)^{2}+
u\left(i\eta\right)
\end{equation}
The conditions to Eqs.~(\ref{2}) are 
\begin{equation} 
\label{4}
\frac{\partial\eta}{\partial\tau}\bigg |_{0}=\frac{\partial\eta}{\partial\tau}\bigg |_{\tau_{0}}=0;\hspace{1cm}
\eta(0)=\eta(\tau_{0})=0
\end{equation}
According to Eq.~(\ref{3}), 
\begin{equation} 
\label{5}
\frac{\partial x}{\partial\tau}\bigg |_{0}=\frac{\partial x}{\partial\tau}\bigg |_{\tau_{0}}=-\sqrt{\frac{2|E|}{m}}
\end{equation}
For a smooth potential all velocities should be zero at the ends of a trajectory but in our case $\partial x/\partial\tau$ is finite 
due to jumps in the potential energy at $x=0$ and $x=R$. The differential equations (\ref{2}), with respect to the functions 
$\partial x/\partial\tau$ and $\eta$, depend on three parameters which should be determined from the six conditions (\ref{4}) and 
(\ref{5}). Since the functions $\partial x/\partial\tau$ and $\eta$ are periodic among the six conditions there are only three 
independent ones. 

\begin{figure}[p]
\begin{center}
\vspace{0.5cm}
\leavevmode
\epsfxsize=\hsize
\epsfxsize=8cm
\epsfbox{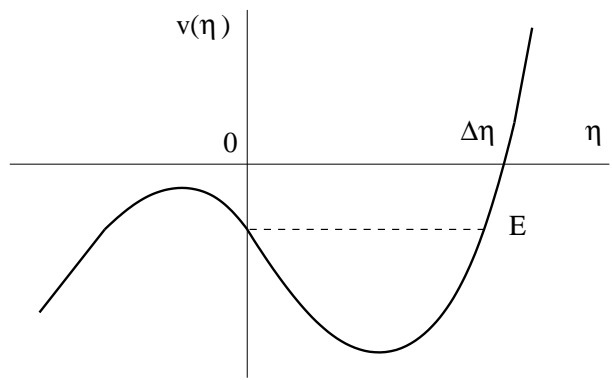}
\vspace{1cm}
\caption{The effective potential for the transverse motion forms the well.}
\label{fig2}
\end{center}
\end{figure}

\begin{figure}[p]
\begin{center}
\vspace{0cm}
\leavevmode
\epsfxsize=\hsize
\epsfxsize=9cm
\epsfbox{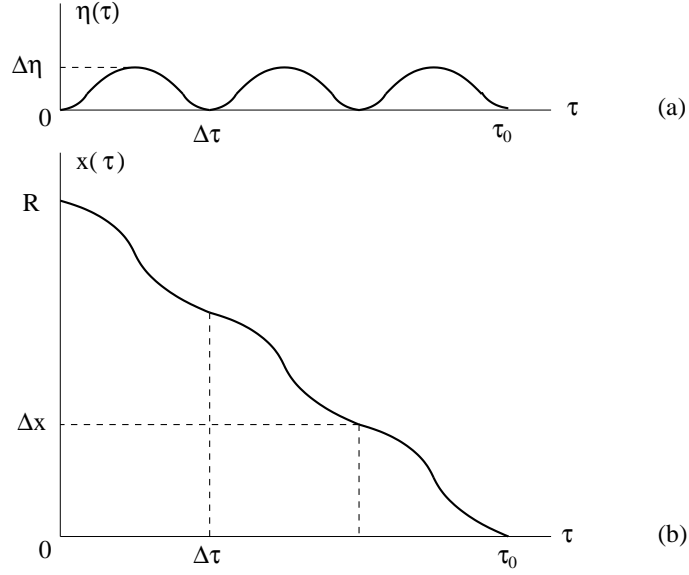}
\vspace{1cm}
\caption{The classical trajectory in imaginary time. It is chosen $N=3$. (a) The transverse component ($y=i\eta$). (b) The motion in the 
direction of tunneling.}
\label{fig3}
\end{center}
\end{figure}

The probability of tunneling $w\sim\exp(-A)$ from $x=0$ to $x=R$ in Fig.~\ref{fig1} is expressed through the Euclidean action 
\cite{COLEMAN1,COLEMAN2,SCHMID1,SCHMID2,GESH,BLATT,GOROKH,MELN1,MELN2,MELN3} 
\begin{equation} 
\label{6}
A=\frac{2}{\hbar}\int^{\tau_{0}}_{0}d\tau\left[\frac{m}{2}\left(\frac{\partial x}{\partial\tau}\right)^{2}-
\frac{m}{2}\left(\frac{\partial\eta}{\partial\tau}\right)^{2}+m\omega_{c}\eta\hspace{0.1cm}\frac{\partial x}{\partial\tau}+
u(i\eta)-E\right]
\end{equation}
Without a magnetic field $\eta =0$ and the action (\ref{6}) coincides with the WKB expression. With the solution of the first equation 
(\ref{2}), $\partial x/\partial\tau =-\omega_{c}\left(\eta +\eta_{0}\right)$, the action (\ref{6}) reads 
\begin{equation} 
\label{7}
A=\frac{2}{\hbar}\int^{\tau_{0}}_{0}d\tau\left[-\frac{m}{2}\left(\frac{\partial\eta}{\partial\tau}\right)^{2}+v(\eta)-E-
m\omega_{c}\eta_{0}\hspace{0.1cm}\frac{\partial x}{\partial\tau}\right]
\end{equation}
By means of Eq.~(\ref{3}) the expression (\ref{7}) takes the form
\begin{equation} 
\label{8}
A=A_{WKB}-\frac{4}{\hbar}\int^{\tau_{0}}_{0}d\tau\left[E-v\left(\eta\right)\right]
\end{equation}
where $A_{WKB}=2\sqrt{2m|E|}R/\hbar$ comes from the last term in Eq.~(\ref{7}) and it is the WKB action related to one-dimensional 
tunneling across a rectangular barrier of the length $R$ with the energy $|E|$ below the barrier top. $A_{WKB}$ in Eq.~(\ref{8}) is generic
with the conventional under-barrier action in a multi-dimensional case \cite{COLEMAN1,COLEMAN2,SCHMID1,SCHMID2}. The second (negative) term
in Eq.~(\ref{8}) is solely due to the magnetic field and corresponds to the transverse motion. The total energy now has the form
\begin{equation} 
\label{9}
E=\frac{m}{2}\left(\frac{\partial\eta}{\partial\tau}\right)^{2}+v(\eta)
\end{equation}
In contrast to the $x$-motion, kinetic energy in the transverse channel does not change sign compared to the physical trajectory 
$\left(\partial y/\partial t\right)^{2}=\left(\partial\eta/\partial\tau\right)^{2}$. Therefore, the transverse motion occurs at the region
where $v(\eta)<E$ and the second term in Eq.~(\ref{8}) is negative. With the expression (\ref{9}) one can describe an oscillatory motion in
the potential $v(\eta)$. We specify a shape of $v(\eta)$ as in Fig.~\ref{fig2} where $v(\Delta\eta)=E$. The trajectory $\eta(\tau)$ is 
drawn in Fig.~\ref{fig3}(a) where the period $\Delta\tau$, according to Eq.~(\ref{9}), is
\begin{equation} 
\label{10}
\Delta\tau =\sqrt{2m}\int^{\Delta\eta}_{0}\frac{d\eta}{\sqrt{E-v(\eta)}}
\end{equation}
The trajectory $x(\tau)$ is shown in Fig.~\ref{fig3}(b). Each cycle of $\eta(\tau)$ in the potential well in Fig.~\ref{fig2} results in the
translation of $x(\tau)$ by $\Delta x$ determined by
\begin{equation} 
\label{11}
\Delta x=\omega_{c}\sqrt{2m}\int^{\Delta\eta}_{0}d\eta\frac{\eta_{0}+\eta}{\sqrt{E-v(\eta)}}
\end{equation}
The trajectory in Fig.~\ref{fig3} looks qualitatively similar as one related to tunneling through a barrier slightly violated by impurities 
in a strong magnetic field \cite{GESH,BLATT}. The difference is that the oscillations in Fig.~\ref{fig3} have an intrinsic nature but in 
\cite{GESH,BLATT} they are determined by distributed impurities. 

This method of trajectories is applicable when the distance between the wells is $R=N\Delta x$ (and also $\tau_{0}=N\Delta\tau$), since at 
the ends ($x=0$ and $x=R$) the condition $\eta =0$ should hold. Fig.~\ref{3} is plotted for $N=3$. By means of 
(\ref{9}), the last term in Eq.~(\ref{8}) can be written as $(-N\Delta A)$ where 
\begin{equation} 
\label{12}
\Delta A=\frac{4\sqrt{2m}}{\hbar}\int^{\Delta\eta}_{0}d\eta\sqrt{E-v(\eta)}
\end{equation}
and the action (\ref{8}) takes the form
\begin{equation} 
\label{13}
A=A_{WKB}-N\Delta A=\left(\frac{2\sqrt{2m|E|}}{\hbar}-\frac{\Delta A}{\Delta x}\right)R
\end{equation}
At a small magnetic field $E-v(i\eta)\simeq -u(i\eta)$ and $\Delta x\sim\ln(1/H)$ is logarithmically big. Therefore, at $H\rightarrow 0$ 
the action (\ref{13}) turns to its conventional limit  $A_{WKB}$.

\begin{figure}[p]
\begin{center}
\vspace{0cm}
\leavevmode
\epsfxsize=\hsize
\epsfxsize=8cm
\epsfbox{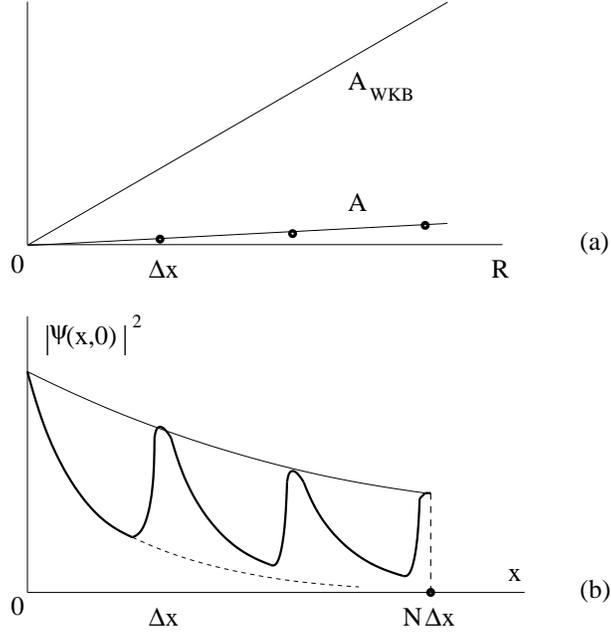}
\vspace{1cm}
\caption{$H<H_{R}$; (a) The action $A$ and its WKB part as a function of distance $R=N\Delta x$, shown by dots, between the two wells. 
(b) Oscillations in the wave function when $R=N\Delta x$ ($N=3$). The dashed curve shows the WKB-like dependence in the absence of 
the magnetic field.}
\label{fig4}
\end{center}
\end{figure}

The action (\ref{13}) and its WKB part are shown in Fig.~\ref{fig4}(a) at the points $R=N\Delta x$ where they have only sense. If $R$ is 
not $N\Delta x$ another method of calculation is required. The action $A$ is substantially reduced compared to its WKB part $A_{WKB}$. When 
the magnetic field is close to the certain value $H_{R}$ the action (\ref{13}) is $A\sim (H_{R}-H)$ and may turn to zero. This means that 
the tunneling probability $w(H)$ becomes not exponentially small at $H=H_{R}$. As mentioned above, the phenomenon, when $A\rightarrow 0$, 
is called Euclidean resonance. 

\begin{figure}[p]
\begin{center}
\vspace{1cm}
\leavevmode
\epsfxsize=\hsize
\epsfxsize=7cm
\epsfbox{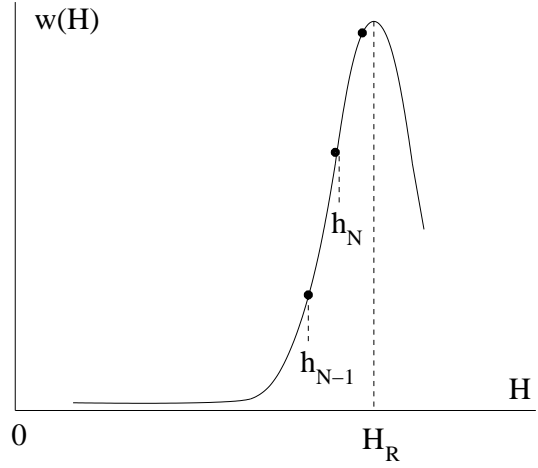}
\vspace{1cm}
\caption{The probability of tunneling as a function of magnetic field has the peak at $H_{R}$ (Euclidean resonance). $h_{N}$ correspond to 
the condition $R=N\Delta x$.} 
\label{fig5}
\end{center}
\end{figure}

The method of trajectory used corresponds to one-instanton approach and it holds at $H<H_{R}$ when $\exp(-A)$ is small. At $H>H_{R}$ the 
exponent $\exp(-A)$ is not small and one should apply a multi-instanton approach which accounts all powers of the exponent. This is a matter 
of a further study. One has to expect that at $H>H_{R}$ the tunneling probability $w(H)$ also decays in a manner as at $H<H_{R}$  having the 
peak at $H=H_{R}$ plotted in Fig.~\ref{fig5}. The peak width, as one can show, is roughly $H_{R}/A_{WKB}$.  The condition $R=N\Delta x(H)$, 
when the above method of trajectories is applicable, holds at certain values of the magnetic field $H=h_{N}$ shown in Fig.~\ref{fig5}. The 
relation 
\begin{equation}
\label{14}
w\left(h_{N-1}\right)=w\left(h_{N}\right)\exp\left(-\Delta A\right)
\end{equation}
says that always exists some $h_{N}$ for which $w(h_{N})$ is not smaller than $\exp(-\Delta A)$. 

The oscillatory structure of the trajectory in the $x$-direction indicates analogous oscillations in the wave function. At $H<H_{R}$ and 
$R=N\Delta x$ the function $|\psi (x,0)|^{2}$ is drawn in Fig.~\ref{fig4}(b) where $|\psi(N\Delta x,0)/\psi(0,0)|^{2}\sim\exp(-A)$. 
Fig.~\ref{fig4}(b) is correct only when the terminal point is $R=N\Delta x$. Fig.~\ref{fig4}(b) does not serve for $H\neq h_{N}$ by a 
simple shift of the terminal point away from $N\Delta x$. 

\begin{figure}[p]
\begin{center}
\vspace{1cm}
\leavevmode
\epsfxsize=\hsize
\epsfxsize=7cm
\epsfbox{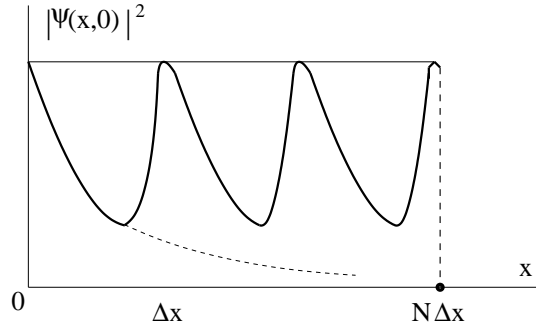}
\vspace{1cm}
\caption{The case of Euclidean resonance $H=H_{R}$. The total distance is $R=N\Delta x$ ($N=3$). The dashed curve represents the WKB-like dependence.}
\label{fig6}
\end{center}
\end{figure}

Under the condition of Euclidean resonance $H=H_{R}$ the function $|\psi(x,0)|^{2}$ in Fig.~\ref{fig6} is periodic at 
least with the exponential accuracy provided by the method of trajectories. This means that $|\psi(x,0)|^{2}$, besides the periodic 
oscillations, may have a power law decay. The spatial oscillations and not exponentially small tunneling through a long barrier draw the 
analogy with motion in a conventional narrow band periodic structure. Like this motion, in our case there is a long distance under-barrier 
coherence set by the magnetic field. One can interpret this as formation of a periodic variable gauge potential. 

An interference of various paths in the problem of localization in disordered systems \cite{MOTT} does not depend on a particular shape of
an impurity potential and is determined by a mean free path. In contrast to this, the under-barrier coherence strongly depends on a 
potential shape since $v(\eta)$ should have a well as in Fig.~\ref{fig2}. This rule provides a choice of $u(y)$. An arbitrary $u(y)$, for 
example $u(y)=u_{0}y^{2}/a^{2}$, does not result in that well but the potential $u(y)=u_{0}\left(y^{2}/a^{2}+y^{4}/a^{4}\right)$ does. 
Analogously, a pure harmonic potential $u(y)=u_{0}\left(1-\cos y/a\right)$ is not suitable and should be supplemented by a double harmonic 
at least. 

Below we consider, as an example, the double harmonic potential $u(y)=u_{0}\left(1-\cos y/a\right)\left(1-\lambda\cos y/a\right)$ which
is created between two quantum dots on a surface with a perpendicular magnetic field. It is sufficient to have this analytical form only
at the length $|y|\lesssim\Delta\eta$. The dots are separated by the distance $R$. Sizes of the dots determine the distance between 
their discrete energy levels which is supposed to be $|E|\simeq 0.01~{\rm eV}$ \cite{WIEL}. The parameters of the double harmonic potential 
are $a=50~\AA$, $\lambda =0.215$, and $u_{0}=1~{\rm eV}$. The spatial period of the potential is $2\pi a=314~\AA$. After calculations one 
can obtain $H_{R}\simeq 10~{\rm Tesla}$, $\Delta x\simeq 120~\AA$, $\Delta\eta\simeq 110~\AA$, and $\Delta A\simeq 12$. At $H=H_{R}$ the 
tunneling probability does not fall exponentially with $R$ which enables to consider tunneling through long barriers, say, $R=1~{\rm cm}$.

The long distance under-barrier coherence can be influenced by dissipation. Dissipation results in a finite width $\delta E$ of energy 
levels inside the wells and also disturbs an under-barrier motion, according to Caldeira and Leggett \cite{LEGGETT}. In the classical 
dynamics dissipation corresponds to the form $m\ddot x+m\gamma\dot x$. Using the theory \cite{LEGGETT}, one can obtain (we omit details) 
the criterion $0.2\gamma\tau_{0}<1$ when dissipation does not influence the frictionless motion. Since $\delta E\sim\hbar\gamma$ this 
criterion is equivalent to $\delta E/E<7.1\lambda/R$ where the parameter $\lambda =\hbar/\sqrt{m|E|}$ can be interpreted as a de Broglie 
wave length. When $\delta E/E\sim 0.1$ \cite{WIEL} one can estimate $R<1000~\AA$. For a bigger $R$ dissipation modifies results and this is 
a matter of a further study. A non-homogeneity $\delta u(x)$ (including applied voltage) of a barrier in the $x$-direction does not 
violate the above results as soon as $\delta u(x)<4|E|$. 

In summary, an answer to the question in the beginning of the paper is positive. The nature allows a long distance under-barrier motion 
which is counterintuitive and contrasts to WKB-like tunneling. An example of this motion, tunneling between two wells in a magnetic field, 
is considered in the paper. In the absence of dissipation the magnetic field sets a long distance coherence under the barrier and a particle 
can tunnel through a long potential barrier as through a conventional narrow band periodic structure. This phenomenon relates to Euclidean 
resonance. 

I thank A. Barone, A. Bezryadin, G. Blatter, M. Gershenson, V. Geshkenbein, L. Ioffe, J. Knight, G. Pepe, A. Ustinov, and R. Webb for 
discussions of related topics.

\newpage

\end{document}